\documentclass[twocolumn,showpacs,amsmath,amssymb,pra]{revtex4}
\usepackage{amssymb,latexsym}
\usepackage{graphicx}
\usepackage{calc}
\usepackage{color}
\renewcommand{\vec}{\textbf}

\newcommand{\be}[1]{\begin{equation}\label{#1}}
\newcommand{\ee}{\end{equation}}
\newcommand{\eq}[1]{Eq.~(\ref{#1})}
\newcommand{\fig}[1]{Fig.~\ref{#1}}

\begin{document}
\title{Rydberg Borromean Trimers}

\author{Ivan C. H. Liu}

\author{Jovica Stanojevic}
\author{Jan M. Rost}
\affiliation{%
Max Planck Institute for the Physics of Complex Systems,
N\"{o}thnitzer Strasse 38, D-01187 Dresden, Germany
}%

\begin{abstract}
 A Rydberg and
a ground-state atom can form  ultralong range diatomic molecules provided the interaction between the ground-state atom and the Rydberg electron is attractive [C. H. Greene, et al., 
Phys. Rev. Lett. 85, 2458 (2000)]. A repulsive interaction does not support bound states. However, as we will show,
adding a second ground-state atom, a bound triatomic molecule becomes possible
 constituting a 
Borromean Rydberg system.
\end{abstract}

\date{\today, version 1.6}

\maketitle
With the advent of ultracold atomic physics new forms of atomic 
systems become possible. The trilobite, arguably one of the most exotic 
molecules, is  a ground-state atom bound to a Rydberg atom by the 
polarization potential of the latter. The existence of this extremely 
long range molecule was proposed, along with its name, by Greene 
\textit{et al}. \cite{Greene00}.  This idea generated considerable interest
\cite{Gra02,Chi02,Ham02,Gre06,Les06}, although a clear
experimental verification has not been achieved so far but may be possible
within a BEC making use of its density orders of magnitude larger
compared to an ultracold gas \cite{Hei07}.
Moreover, previously unexplained satellite peaks in thermal 
spectra of molecules could be traced to the formation of such long 
molecules \cite{Gre06}. The basis of the molecular bond in such systems is a negative
scattering length leading to an attractive atom-Rydberg atom potential which
is in addition modulated by the electronic Rydberg 
wavefunction.
An extension of the trilobite to 
polyatomic Rydberg molecules involving several ground-state atoms and 
one Rydberg atom was investigated in \cite{Liu06}. There, it was 
found that certain geometric arrangements have adiabatic molecular 
potentials which are many times deeper than in the diatomic case.

However, a Rydberg molecule can even form with a positive scattering length, i.e., repulsive interaction, as we
will show in the following.  While a diatomic molecule is indeed not
possible under repulsive interaction, a triatomic molecule can be
formed which we call Borromean.  The name comes from a similar
phenomenon in nuclear physics \cite{Zhu93}.  The essence of a
Borromean system is the existence of a three-body bound state although
none of the two-body subsystems are bound.  For the origin of
``Borromean rings'' see \cite{Liv96}.

In the following, we will show that the building principles of such 
ultralong range molecules allow the formation of a Borromean 
trimer using neon as an example.
This is a molecule formed out of three linearly aligned 
neon atoms where the one in the middle is Rydberg excited. 
The mutual interaction of two neon atoms is repulsive and none of 
the atomic pairs has a bound state. Nevertheless, the triatomic 
system can have a bound state. We will use atomic units unless stated otherwise.

Since the Rydberg wavefunction does not vary over the extension of a
ground-state electronic wavefunction, the interaction can be described
by a Fermi-pseudopotential \cite{Fermi34}, which amounts to a contact
interaction 
\be{fermipotential} 
V_{F}(\vec r;\vec r_{1}) = L_{1} \delta(\vec
r -\vec r_{1})\,, 
\ee 
where $\vec r$ is the coordinate of the Rydberg
electron and $\vec r_{1}$ is the position of the ground-state atom $A$ measured 
from the Rydberg atom.
Furthermore, $L_1=- 2\pi \tan\delta_{0}/k(r_{1})$ is the \textit{energy-dependent}
scattering length of the collision system $e^{-}+A$  containing
the $s$-wave phase shift $\delta_0$  and   the local momentum $k(r_1)$ of the Rydberg electron,
related to the distance $r_1$ through the energy conservation
$k(r_1)^2/2=-1/2n^2+1/R.$
In the low-energy limit, the modified effective range theory \cite{Omalley61}
 expresses $L_1$
analytically in terms of the zero-energy scattering length $L_0$ and the polarizability of 
the atom $\alpha$, 
\be{slength}
\frac{L_1}{2\pi}=L_0 + \frac{\pi}{3}\alpha k(r_1).
\ee
For impact energy larger than 0.003 eV in the $e^-+\textrm{Ne}$ system,
\textit{ab initio} phase shift data \cite{Sah90} allow one
to connect the scattering length of Eq.~\eqref{slength} to higher energies
accurately, using the values $L_0=2.613$ a.u. and $\alpha=0.2218$ a.u. from \cite{Sah90}.

For $N$ ground-state atoms at positions $\vec r_{i}$, the full 
electronic Hamiltonian reads
\be{hamil}
h_{N} = \frac{p^{2}}{2}+V_{c}(r)+\sum_{i=1}^{N}L_{i}\delta(\vec r-\vec 
r_{i})\,,
\ee
where $V_{c}(r)$ is the Coulomb interaction of the Rydberg electron 
with its mother ion. 

The eigenenergy of a dimer $h_{2}$ may be obtained by diagonalizing the interaction 
$V_{F}$ within a degenerate manifold of hydrogenic eigenfunctions $\Phi_{nlm}$ of 
principal quantum number $n$. This leads to the (unnormalized) solution
\be{eigenf2}
\psi_{n}(\vec r;\vec r_{1}) = \sum_{l\ge l_{\mathrm 
min},m}\Phi^{*}_{nlm}(\vec r_{1})\Phi_{nlm}(\vec r)\,.
\ee
Low-$l$ states 
may have different energies due to large quantum defects, so they 
need to be excluded from the basis 
formed with states of Coulomb energy $-1/2n^{2}$.

\begin{center}
\begin{figure}
\includegraphics[width=\columnwidth*\real{0.85}]{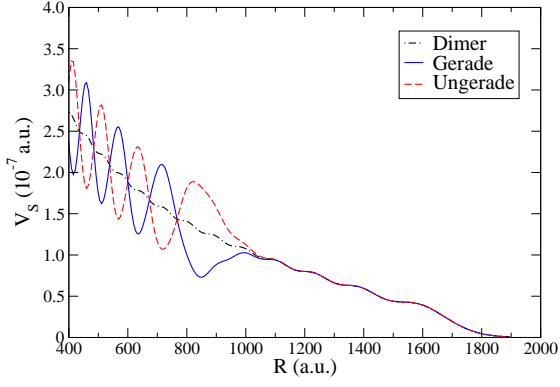}
 \caption{\label{Fi:SStretch} The adiabatic potential \eq{energy2} with $r_{1}=R$ for the ($n=30$) neon Rydberg dimer (dash-dotteded black) and the symmetric stretch cuts 
 \eq{SS} of trimer in solid blue ($+$) and dashed red ($-$).}
\end{figure}
\end{center}

Making use of the addition theorem for 
spherical harmonics ($\hat x$ is a unit vector) \cite{Var88},
we reduce \eq{eigenf2} to
\be{eigenf22}
\psi_{n}(\vec r;\vec r_{1}) = \sum_{l\ge l_{\mathrm 
min}}\phi_{nl}(r_{1})\phi_{nl}(r)\frac{2l+1}{4\pi}P_{l}(\hat 
r_{1}\hat r)\,,
\ee
with the radial hydrogenic eigenfunctions $\phi_{nl}$.
It is convenient to specify the overlap intergral
$N_{12}\equiv \langle\psi_{n}(\vec r;\vec r_{1})|\psi_{n}(\vec r;\vec 
r_{2})\rangle_{\vec r}$ which yields
\be{nab}
N_{12} = 
\sum_{l}\phi_{nl}(r_{2})\phi_{nl}(r_{1})\frac{2l+1}{4\pi}P_{l}(\hat 
r_{2}\hat r_{1})\,.
\ee
Clearly, the normalized eigenfunction \eq{eigenf2}  reads 
$\Psi_{n}(\vec r;\vec r_{1}) = \psi_{n}(\vec r;\vec 
r_{1})/N_{11}^{1/2}$. 
For future reference we define
\be{definitions}
N^{\pm}_{11} = (\psi_{n}(\vec r_{1};\vec 
r_{1})\pm\psi_{n}(\vec r_{1};-\vec r_{1})  )/2\,.
\ee
The relation Eq.~\eqref{nab} allows one to write the dimer energy 
$
E_{2}(r_{1})=\langle\Psi_{n}|h_{2}(r_{1})|\Psi_{n}\rangle+1/2n^{2}$
in the compact form
\be{energy2}
E_{2}(r_{1}) = L_{1}N_{11}\,.
\ee
The repulsive interaction $V_{F}$ for the neon Rydberg 
dimer Ne+Ne$^{*}$ $E_{2}$ is
shown as a dash-dotted line in \fig{Fi:SStretch}. Clearly, this system does 
not have bound states.

However, we will show that adding another neon ground-state atom 
allows for bound states of the neon trimer with Hamiltonian
\be{trimer}
h_{3} = \frac{p^{2}}{2}+V_{c}(r)+L_{1}\delta(\vec r-\vec 
r_{1})+ L_{2}\delta(\vec r-\vec r_{2})\,,
\ee
where $L_{i}>0$.
The following basis functions are defined which are symmetry-adapted linear 
combinations of dimer eigenfunctions with gerade/ungerade ($\pm$) symmetry 
(since the two neon atoms 1 and 2 are identical),
\be{eigenf3}
\Psi_{\pm}(\vec r;\vec r_{1},\vec r_{2}) = [\Psi_{n}(\vec r;\vec 
r_{1})\pm\Psi_{n}(\vec r;\vec r_{2})]/N^{1/2}_{\pm} \,,
\ee
with the normalization factors
\be{}
N_{\pm}=2\pm2\frac{N_{12}}{\sqrt{N_{11}N_{22}}}\,.
\ee
In this set of basis functions, the Hamiltonian $h_3$ forms a
$2\times2$ matrix
with diagonal terms
$E^{\pm}_{\mathrm{dia}} = 
\langle\Psi_{\pm}|h_{3}|\Psi_{\pm}\rangle+1/2n^{2}$
of the simple form
\be{energy3}
E_{\mathrm{dia}}^{\pm}(\vec r_{1},\vec 
r_{2})=\frac{L_{1}N_{11}+L_{2}N_{22}}{4}N_{\pm}\,.
\ee
The off-diagonal terms 
$E_{\mathrm{off}} = \langle\Psi_{\pm}|h_{3}|\Psi_{\mp}\rangle
+1/2n^{2}$
read
\be{energy3-off}
E_{\mathrm{off}}(\vec r_{1},\vec 
r_{2})=\frac{L_{1}N_{11}+L_{2}N_{22}}{4}\sqrt{N_{+} N_{-}}\,.
\ee

A symmetric cut through this potential for $\vec r_{2}=-\vec 
r_{1}\equiv \vec R$
is shown in \fig{Fi:SStretch}.  This cut has the appealing analytical form
\be{SS}
E_{\mathrm{dia}}^{\pm}(\vec r_{1},-\vec r_{1}) =2L_{1}N^{\pm}_{11}\,. 
\ee
Hence, depending on gerade/ungerade symmetry only even/odd components
$l$ in the sum of \eq{nab} contribute. 
The symmetry-induced separation of basis functions with even and odd $l$
results in a more pronounced undulation of the corresponding potentials 
(as can be seen in \fig{Fi:SStretch}),
since there is less intereference 
from wavefunctions with different nodal structure.
For the full solution the $2\times2$ potential matrix 
can be diagonalized whose eigenvalues give the adiabatic potentials in the analytical form
\be{matrix}
\begin{split}
E^{\mathrm{u/d}} (\vec r_{1},\vec r_{2})
&= \frac{L_1 N_{11}+L_2 N_{22}}{2} \\
& \pm \frac{\sqrt{ (L_1 N_{11} - L_2 N_{22})^2 + 4 L_1 L_2 N_{12}^2 }}{2}\,.
\end{split}
\ee

To determine if the neon Rydberg trimer can support bound states we 
make a normal mode analysis on the electronic surface from \eq{matrix},
to which the outermost minimum belongs.
To this end it is useful to 
define symmetry adopted Jacobi coordinates for the nuclear motion,
\be{coordinates}
    \vec r_{A} =(\vec r_{1}+\vec r_{2})/2,\,\,\,\,\,\,\,
    \vec R =(\vec r_{1}-\vec r_{2})/2\,.
\ee
    In these coordinates, the nuclear Hamiltonian for total 
angular momentum $L=0$ is given by
\be{hamil}
H = \frac{1}{4}\frac{P^{2}}{m}+\frac{3}{4}\frac{p_{A}^{2}}{m}+E^{\mathrm{u/d}}( 
\vec r_{A} + R\hat z,\vec r_{A}-R\hat z)\,,
\ee
where $m$ is the mass of a neon atom, and 
we have chosen the coordinates such that $\vec R = R\hat z$ 
along the body fixed $z$-axis. We may re-express the potential energy 
surface as $E^{\mathrm{u/d}}={\cal E}^{\mathrm{u/d}}(R,z,\rho)$. 

In the following, we quantize the trimer with $n=30$ in a separable approximation
 closely related to the normal mode analysis for molecules.
With the condition
$d{\cal E}^{\mathrm{d}}/dR=d{\cal E}^{\mathrm{d}}/d\rho=d{\cal E}^{\mathrm{d}}/dz=0$  we determine the 
equilibrium points $(R,\rho,z)=(R_{i},0,0)$ of the potential. 
Figure \ref{Fi:SStretch} reveals many minima in the symmetric stretch (SS) 
coordinate $R$; we will quantize the outermost minimum at 
$R_{0}=850\,$au.  As can be seen on the same figure,   
 tunneling of the quantized states in the $R$ coordinate  
 depends sensitively on the form of the potential. Hence, instead
 of a formal harmonic normal mode expansion about the  equilibrium point, we simply split the potential in 
 several contributions, leaving only one coordinate as a variable in 
 each term while fixing the others at their equilibrium values. This allows for an explicit numerical solution in each degree of freedom under the total potential
 \be{expansion}
 {\cal E}^{0}(R,z,\rho) =  V_{S}(R)+V_{A}(z)+V_{B}(\rho)-2V_{0}\,,
 \ee
 with
 \begin{eqnarray}
 \label{expansion2}
V_{0}&=&{\cal E^{\mathrm{d}}}(R_{0},0,0)\\\nonumber
V_{A}(z)&=&{\cal E^{\mathrm{d}}}(R_{0},0,z)\\\nonumber
V_{S}(R)&=&E^{\mathrm +}_{\mathrm{dia}}(R\hat z,-R\hat z)\\\nonumber
V_{B}(\rho)&=& E^{\mathrm +}_{\mathrm{dia}}
 (\rho \hat{x} + R_0\hat{z}, \rho \hat{x} -R_0\hat{z})\,.
 \end{eqnarray}
 The modes correspond to molecular types of normal mode vibrations along the SS (in $R$), the bending (in $\rho$) and the asymmetric stretch (AS) (in $z$) with the product wavefunction
\be{normalmode}
\Psi^{0}(R,z,\rho) = \chi(R)\phi(z)\xi(\rho)\,,
 \ee
 where
 \be{eqas}
 \left(-\frac{1}{4m}\frac{d^{2}}{dR^{2}}+V_{S}(R)\right)\chi(R) = 
 E_{S}\chi(R)
 \ee
 and similarly for the other coordinates.
 
\begin{center}
\begin{figure}
\includegraphics[width=\columnwidth*\real{0.85}]{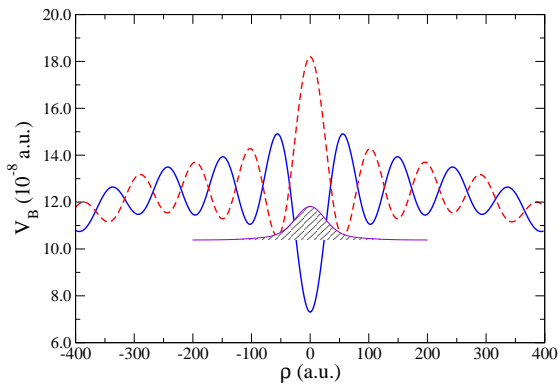}
 \caption{\label{Fi:Bending} The adiabatic potential $V_{B}(\rho)$ (see \eq{expansion2}) at the outermost equilibrium point $R_{0}$ as a function of the bending coordinate $\rho$ for  the ($n=30$) neon Rydberg trimer. The solid blue and the dashed red correspond to
 $E^{+}_{\mathrm{dia}}$ and $E^{-}_{\mathrm{dia}}$
 in \eq{energy3} respectively.
The shaded area shows the quantized state.}
\end{figure}
\end{center}

\begin{center}
\begin{figure}
\includegraphics[width=\columnwidth*\real{0.85}]{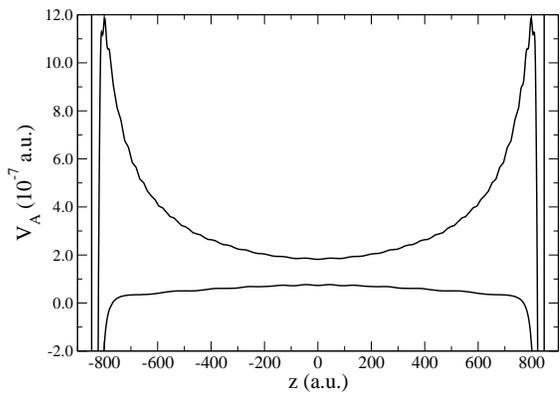}
 \caption{\label{Fi:Astretch} The adiabatic potential $V_{A}(z)$ (see \eq{expansion2}) at the outermost equilibrium point $R_{0}$ as a function of the asymmetric stretch coordinate $z$ for  the ($n=30$) neon Rydberg trimer as well as the corresponding potential for the ${\cal E^{\mathrm{u}}}$ potential (upper curve).}
\end{figure}
\end{center}


The quantized bending motion has a zero-point energy at
$E_{B}=1.04 \times 10^{-7}\,$au as shown in \fig{Fi:Bending}. 
The SS potential shown in \fig{Fi:SStretch} supports a ground state with energy $E_{S}=8.11\times 10^{-8}\,$au and an excited state.
Both of them are in principle resonances which decay through tunneling in the separable approximation. However, the ground state
is practically stable.  Last, we come to the AS which has near equilibrium ($z=0$) the form of an inverted oscillator (\fig{Fi:Astretch}). The downwardly-bent curve falls on the left and right into deep but narrow wells, which are the normal Ne$_{2}^{+}$ electronic ground-state potentials, to which the asymptotic polarization potential
$-\alpha/2r_i^4$ of the Rydberg excited level is smoothly connected. 
For even shorter distances, the \textit{ab initio} data for the Ne$_{2}^{+}$ potential
from Ref.~\cite{Coh74} are used.
The potential landscape is similar to those of triatomic ABA molecules for which so called hyperspherical resonances exist \cite{Man84}. The corresponding  quasibound resonant eigenfunctions are highly excited with many nodes along the AS and the resonant energy $E_{A}$ almost touches the top of the barrier, i.e., the equilibrium point $E_{A}\approx V_{0}= 7.30\times 10^{-8} $au. To a good approximation the lifetime of the resonance
can be estimated by the curvature $k_{A}$ of the potential at $V_{0}$, $\tau_A\approx
2\pi\omega_{A}^{-1}=2\pi\sqrt{\mu_A/k_A} = 57$ ns with $\mu_A=2m/3$. Intuitively, this curvature determines the time scale on which probability density ``slides down'' the inverted oscillator potential towards the dimer diatomic decay channel. To summarize, the total ground state energy of the Borromean trimer measured from 
the Rydberg energy $-1/2n^{2}$ is given by
\be{energy_{tot}}
E_{0}= E_{S}+E_{A}+2E_{B}-3V_{0} = 1.43\times 10^{-7}\, \mbox{au}.
\ee
This means a blue shift of 940 MHz from the Rydberg excitation line and the lifetime is limited by the decay of the AS mode to about $\tau \approx 57$ ns.
Due to the cylindrical symmetry of the collinear trimer in equilibrium, the bending motion in $(\rho,\phi)$ is doubly degenerate, hence it contributes $2E_{B}$.

\begin{figure}
\centering
\includegraphics[width=\columnwidth*\real{0.7}]{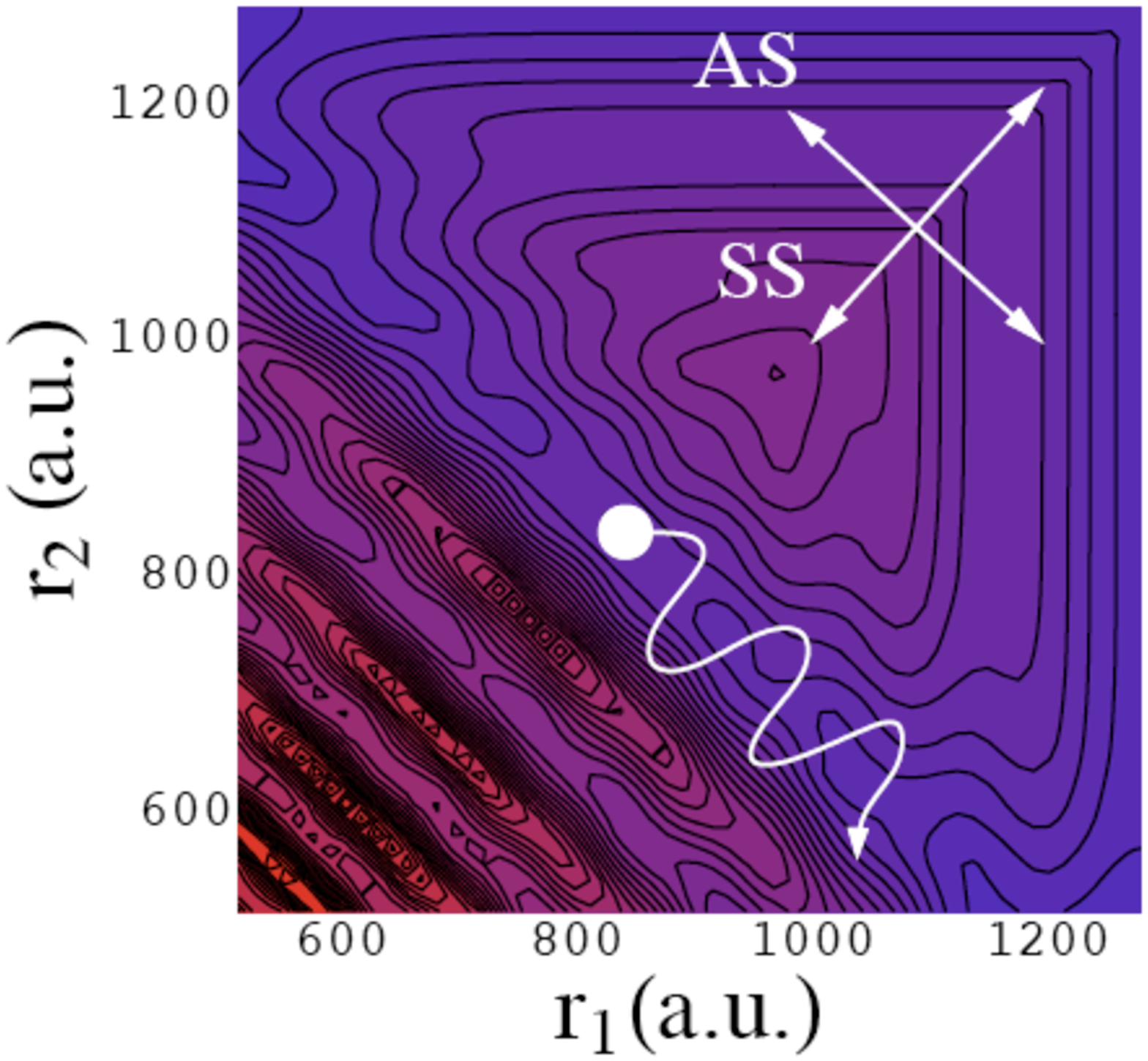}
\includegraphics[width=\columnwidth*\real{0.14}]{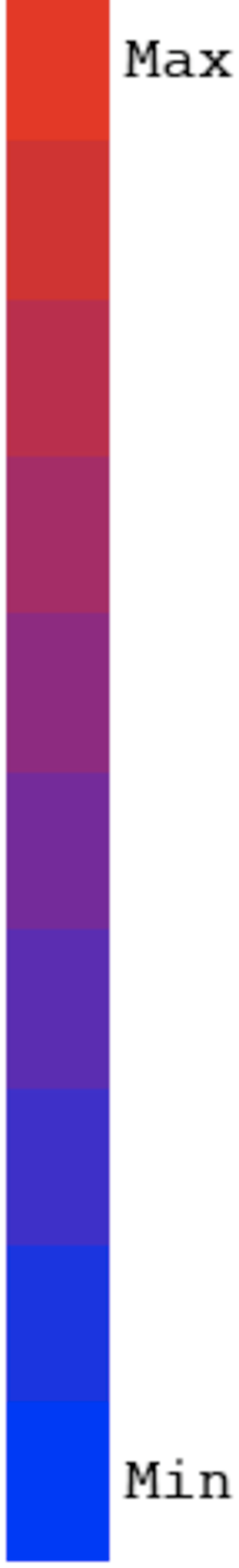}
 \caption{\label{Fi:contours} The adiabatic potential for the lower
 surface of the ($n=30$) neon Rydberg trimer in collinear configuration. 
 Symmetric (SS) and asymmetric stretch (AS) modes are indicated, as well as the dynamics of the trimer.}
\end{figure}

For $n=30$, the bending can be neglected due to the localized motion about $\rho=0$
and the long bond-length.
Figure \ref{Fi:contours} gives an overview of the dynamics of the trimer in the plane of the SS and AS coordinate (in collinear geometry with $\rho = 0$).
The system rolls down from the top of the saddle point through a potential valley
into the two-body region.

The lifetime of the Rydberg Borromean trimer can be optimized by varying 
the Rydberg excitation  $n$:
Higher $n$ leads to shallower potentials which enlarges the lifetime in the AS but shortens it in the SS and the bending
 due to a stronger tunneling. 
 Hence, there exists some $n$ with the maximum $\tau$.
To determine its value, we first obtained a general expression for the curvature 
of the saddle in the AS, which
gives an estimate for the decaying time
$\tau_A=n^4\sqrt{2\pi^3\mu_A/L_0}$.
This is obtained by approximating the shape of the saddle with a parabola of height
$L_0/\pi n^4$ and width $2n^2$.
We then numerically calculate the tunneling times $\tau_S$ and $\tau_B$
in the SS and the bending by solving the corresponding Schr{\"o}dinger
equations and matching the asymptotic phases, for
increasing $n$ at the outer-most minimum until the tunneling in the bending becomes too large.
For Ne$^*_3$, it was found that this optimal value occurs at about $n=55$, where
the overall lifetime is as large as
$\tau=(1/\tau_S+2/\tau_B+1/\tau_A)^{-1}=0.46\mu$s.


We note in passing that a similar analysis of the Rydberg trimer under negative scattering length for $^{87}$Rb produces a bound state, \emph{red shifted} by 
-12.7 GHz  with respect to the Rydberg line, with the corresponding potential
minimum $V_0=-14.3$ GHz at $(865,0,0)$.

One may question the validity of the adiabatic approximation in the presence of a series of conical intersections as obvious from the crossings in the diagonal part of the hamiltonian matrix \eq{SS}
and \fig{Fi:SStretch}. However, the vibrational ground-state wavefunction $\chi_{0}$ is strongly localized about $R_{0}$, suppressing the effect of coupling even at the closest conical intersection.  

Future work will consider the effect of a series of conical sections in a situation where they 
are dynamically active, i.e.,  not sitting on a decaying background potential which renders the adiabatic wavefunction exponentially small under a larger barrier at the location of the intersections. 

To summarize, we have demonstrated that a Rydberg atom may form with two ground-state atoms an ultralong range Borromean molecule with finite but long lifetime for positive scattering length of the Rydberg electron-ground-state-atom interaction. We call it Borromean since the
corresponding diatomic molecule Rydberg and ground-state atom, do not have a bound state or long lived resonance.

 We thank  F\'{e}d\'{e}ric Merkt for drawing our attention to the problem of the repulsive Rydberg dimer.
 ICHL thanks the International Max Planck Research School ``Dynamical Processes in
 Atoms, Molecules and Solids'' for financial support.
 

\end{document}